\documentclass[%
superscriptaddress,
 amsmath,amssymb,
 aps,
 prstab, twocolumn,
]{revtex4-1}
\usepackage{graphicx}
\usepackage[utf8]{inputenc}
\usepackage{color}
\usepackage{amsmath}
\usepackage{color}
\usepackage{amssymb}
\usepackage{amsmath}
\usepackage{tabularx}
\usepackage{booktabs}
\usepackage{multirow}
\usepackage[colorlinks, citecolor=blue, urlcolor=black, bookmarks=false, linkcolor=blue, hypertexnames=true]{hyperref}

\def\ss#1{\textcolor{black}{#1}}
\def\ssrev#1{\textcolor{black}{#1}}

\begin{document}
\title{High-temperature phonon-mediated superconductivity in monolayer Mg$_2$B$_4$C$_2$}


\author{Sobhit Singh}
\email{sobhit.singh@rutgers.edu}
\affiliation{Department of Physics and Astronomy, Rutgers University, Piscataway, New Jersey 08854, USA}

\author{Aldo H. Romero}
\email{Aldo.Romero@mail.wvu.edu}
\affiliation{Department of Physics and Astronomy, West Virginia University, Morgantown, West Virginia, USA}

\author{Jos\'e D. Mella}
\affiliation{Departamento de F\'isica, Facultad de Ciencias F\'isicas y Matem\'aticas, Universidad de Chile, Santiago, Chile}
\affiliation{Departamento de F\'isica, Facultad de Ciencias, Universidad de Chile, Santiago, Chile}

\author{Vitalie Eremeev}
\affiliation{Instituto de Ciencias B\'asicas, Facultad de Ingenier\'ia y Ciencias, Universidad Diego Portales, Santiago, Chile}

\author{Enrique Mu\~noz}
\affiliation{Institute of Physics, Pontificia Universidad Cat\'olica de Chile, Santiago, Chile}

\author{Anastassia N. Alexandrova}
\affiliation{Department of Chemistry and Biochemistry, University of California, Los Angeles, CA 90095, USA}
\affiliation{California NanoSystems Institute, Los Angeles, CA 90095, USA}

\author{Karin M. Rabe}
\affiliation{Department of Physics and Astronomy, Rutgers University, Piscataway, New Jersey 08854, USA}

\author{David Vanderbilt}
\affiliation{Department of Physics and Astronomy, Rutgers University, Piscataway, New Jersey 08854, USA}

\author{Francisco Mu\~noz}
\email{fvmunoz@u.uchile.cl}
\affiliation{Departamento de F\'isica, Facultad de Ciencias, Universidad de Chile, Santiago, Chile}
\affiliation{Center for the Development of Nanoscience and Nanotechnology (CEDENNA), Santiago, Chile}

\begin{abstract}
  A new two-dimensional material -- Mg$_2$B$_4$C$_2$, 
  belonging to the family of the conventional superconductor MgB$_2$, 
  is theoretically predicted to exhibit superconductivity with critical temperature 
  $T{_c}$ estimated in the \ss{47--48\,K range} (predicted using \ssrev{the McMillian-Allen-Dynes formula}) 
  without any tuning of external parameters 
  such as doping, strain, or substrate-induced effects. 
  The origin of such a high intrinsic $T{_c}$ is ascribed to the 
  presence of strong electron-phonon coupling and topological Dirac states 
  (which are absent in MgB$_2$) yielding a large density of states at the Fermi level.
  This material also features a nontrivial electronic band topology exhibiting 
  Dirac points, practically gapless Dirac nodal lines, and topological nontrivial edge states. 
 Consequently, it is a potential candidate for realization of topological superconductivity in 2D. 
 This system is obtained after replacing the chemically active 
  boron layers in MgB$_2$ by chemically inactive boron-carbon layers. 
  Hence, the surfaces of this material are inert. 
  Our calculations confirm the stability of 2D Mg$_2$B$_4$C$_2$. 
  We also find that the key features of this material remain essentially
  unchanged when its thickness is increased by modestly increasing the number 
  of inner MgB$_2$ layers. 
  \end{abstract}

\maketitle

\section*{Introduction}

The discovery of highly crystalline two-dimensional (2D) 
superconductors~\cite{Takashi17, saito2017highly, Jiang14, Kamihara_JACS2008, Brun2016}, 
such as NbSe$_2$ monolayer~\cite{FrindtPRL1972, Xiaoxiang2015, Ugeda2016, LiPRB2019}, 
has provided new possibilities for van der Waals (vdW) heterostructures 
nano-engineering of novel insulator-superconductor interfaces~\cite{Novoselovaac2016} 
and 2D Josephson junctions, without the need of an insulating layer~\cite{yabuki2016s}. 
One main challenging issue in the realization of 2D superconductivity is that 
most of the well-known conventional bulk superconductors 
either do not superconduct or poorly superconduct when 
their dimensions are reduced~\cite{FrindtPRL1972, Xiaoxiang2015, Ugeda2016, xi2016ising, 
BoXu2016, BarreraNatComm2018, LBao2018, Yan_2019, YanLuo2020}. 
%
Although numerous 2D phonon-mediated superconductors have recently been predicted 
from first-principles calculations, the highest predicted intrinsic $T_c$ stayed around 
20\,K~\cite{penev16, Wang18, Lei17, dai12, GaoPRB2017, ZiyangQu2019, LuoYan_PCCP2019, YanLuo2020} 
(19\,K for B$_2$C monolayer~\cite{dai12}, \ss{10.3\,K for B$_2$O monolayer~\cite{YanLuo2020}, }
and 19--25\,K for borophenes~\cite{GaoPRB2017}, to name a few). 
Though in some cases $T_c$ has been enhanced by means of the chemical doping, 
intercalation, strain, and/or substrate proximity 
effects~\cite{Rosner_PRL2002, Pogrebnyakov2004, Weller2005, Gauzzi_PRL2007, 
SaviniPRL2010, QishengWu2016,Zhang2017, ZiyangQu2019, Bekaert2019, YanLuo2020},
 it is essential to discover intrinsic 2D superconductors that exhibit 
high-$T_c$ without any doping or tuning of external parameters 
(here high-$T_c$ does not refer to unconventional superconductivity as in case of 
cuprates or iron-based superconductors~\cite{kruchinin2014physics}). 

Among all the Bardeen--Cooper--Schrieffer (BCS) type conventional 
superconductors, MgB$_2$ stands out with a record $T_c$ of 39\,K, 
the highest reported $T_c$ at zero-pressure~\cite{nagamatsu2001, ChoiNature2002, Choi_PRB2002}.  
Such a high-$T_c$ in MgB$_2$ stems from the strong electron-phonon (el-ph) coupling 
occurring primarily due to the in-plane stretching of B-B bonds ($i.e.$, $E_{2g}$ phonon modes), 
which strongly couple with the self-doped charge carriers from magnesium to 
boron atoms~\cite{Pickett_PRL2001, bohnen2001phonon, ChoiNature2002, Choi_PRB2002, Pogrebnyakov2004}. 
\ssrev{Remarkably, only two ($E_{2g}$) out of a total of nine phonon modes contribute strongly 
to the total el-ph coupling in MgB$_2$~\cite{Pickett_PRL2001, bohnen2001phonon, Iavarone02, 
ChoiNature2002, Choi_PRB2002, Pickett_Nature2002, MAZIN200349, XXXi2008, PICKETT2008, Pickett_design2006}. }
Once the fundamental mechanism of such a high-$T_c$ in bulk MgB$_2$ was understood, 
which by the way was a subject of intense research for over a decade 
period~\cite{Pickett_PRL2001, Yildirim01, Iavarone02, Szab_PRL2001, bohnen2001phonon, 
ChoiNature2002, Choi_PRB2002, Pickett_Nature2002, RosnerPRB2002, MAZIN200349, 
CHOI200366, Pogrebnyakov2004, KORTUS200754, KORTUS200754, XXXi2008, PICKETT2008, Pickett_design2006}, 
researchers started proposing novel ways to augment $T_c$ through 
rational material design approach~\cite{PICKETT1981_MoN, Pickett_PRL2001, 
Rosner_PRL2002, PICKETT_2d_2003, Pickett_design2006, PICKETT2008, BoeriJPCM2021}. 
Pickett and co-workers proposed that one can, in principle, 
achieve a much higher $T_c$ (than 39\,K) by designing a MgB$_2$-like stable 
material which has a similar Fermi surface as in MgB$_2$, and 
in which more than two phonon modes couple to the electronic states 
near the Fermi level, thereby, resulting in a sizable total el-ph 
coupling~\cite{Pickett_design2006, PICKETT2008, Rosner_PRL2002}. 
This idea has been employed for the rational design of new bulk superconductors 
with a good success rate~\cite{verma03, Rosner2002, AnPRB2002, Choi09, Miao16, 
Bersier09, Norman16, stanev2018machine, Zhai19, klintenberg2013, KolmogorovPRL2010, GouPRL2013}.  
\ss{The high-pressure superconductivity observed at 250 K in lanthanum hydride is one 
such example~\cite{LiuPNAS2017, PengPRL2017, LiuPRB2018, Drozdov2019}.}

Despite the large success with the bulk conventional superconductors, 
two-dimensional intrinsic superconductors having a high-$T_c$ remained elusive. 
Notably, various attempts have been made to realize superconductivity in 
the 2D analogues of bulk MgB$_2$~\cite{PICKETT_2d_2003, Naito_2004, Xi_2009_2d, MazinBalatsky2010, jishi2011theoretical, BoXu2016, Bekaert2017, bekaert2017free, Bekaert2019}. 
On the one hand, Xu and Beckman proposed a quasi-2D MgB$_2$ nanosheet with inert surfaces, 
which turns out to be a semiconductor with a bandgap of 0.51\,eV resulting from the 
quantum confinement effects~\cite{BoXu2016}. 
On the other hand, Bekaert et al. reported that a considerably high-$T_c$ of 20\,K 
can be realized in monolayer MgB$_2$ without surface passivation, 
$i.e.$, if only such a material with a highly chemically reactive surface could be made 
~\cite{Bekaert2017, bekaert2017free}. 
\ss{In a recent study, Bekaert et al. theoretically demonstrated that a MgB$_2$ monolayer 
can be stabilized by adding hydrogen adatoms. Interestingly, they find that the hydrogenation 
process leads to a high-$T_c$  of 67\,K, which can be further boosted to over 100\,K by 
means of a biaxial strain on the hydrogenated MgB$_2$ monolayer~\cite{Bekaert2019}.  } 
While an experimental validation of the predicted $T_c$ in monolayer MgB$_2$ is still missing, 
the aforementioned theoretical works markedly enhance our understanding of 
superconductivity in 2D materials.

In this work, we present a novel MgB$_2$-like 2D material -- Mg$_2$B$_4$C$_2$, 
having charge neutral inert surfaces, which is predicted to superconduct at a 
strikingly high-$T_c$ in the \ss{47--48\,K range} (predicted using the 
McMillian-Allen-Dynes theory~\cite{Migdal1958, Eliashberg60, PBAllen1983}), 
which is \ss{among the highest $T_c$ yet reported} for an intrinsic 2D material without any doping, 
strain or substrate-induced effects. 
The main advantageous feature in 2D Mg$_2$B$_4$C$_2$ is the fact that, 
unlike in bulk MgB$_2$, more than two phonon modes strongly couple to the 
electronic states near the Fermi level, thus, resulting in a substantially 
larger el-ph coupling ($\lambda = 1.40$) in monolayer Mg$_2$B$_4$C$_2$ 
compared to the bulk MgB$_2$ ($\lambda_{bulk} = 0.73$~\cite{bohnen2001phonon}, 
and 0.61~\cite{Choi_PRB2002}). \ss{We note that the estimated $\lambda$ in monolayer 
Mg$_2$B$_4$C$_2$ is comparable to the predicted $\lambda$ (=1.46) in hydrogenated 
MgB$_2$ monolayer~\cite{Bekaert2019}.}
Moreover, our calculations reveal nontrivial topological electronic features in 
Mg$_2$B$_4$C$_2$ exhibiting Dirac cones and practically gapless Dirac 
nodal lines \ssrev{at the Fermi level near the corner points of the hexagonal Brillouin zone (BZ), which enhance 
the density of states (DOS) at the Fermi level by almost 30\% compared to that of bulk MgB$_2$,} hence, positively contributing 
towards a higher $T_c$.  
\ss{Although we do not establish any connection between the predicted superconductivity 
 and the nontrivial topological electronic properties in this work, we note 
that the aforementioned two features are often required for realization of topological 
superconductivity~\cite{Kobayashi_PRL2015}, investigation of which is beyond 
the scope of the present work. 
}

\section*{Results}

\begin{figure*}[ht!]
\centering \includegraphics[width=0.9\textwidth]{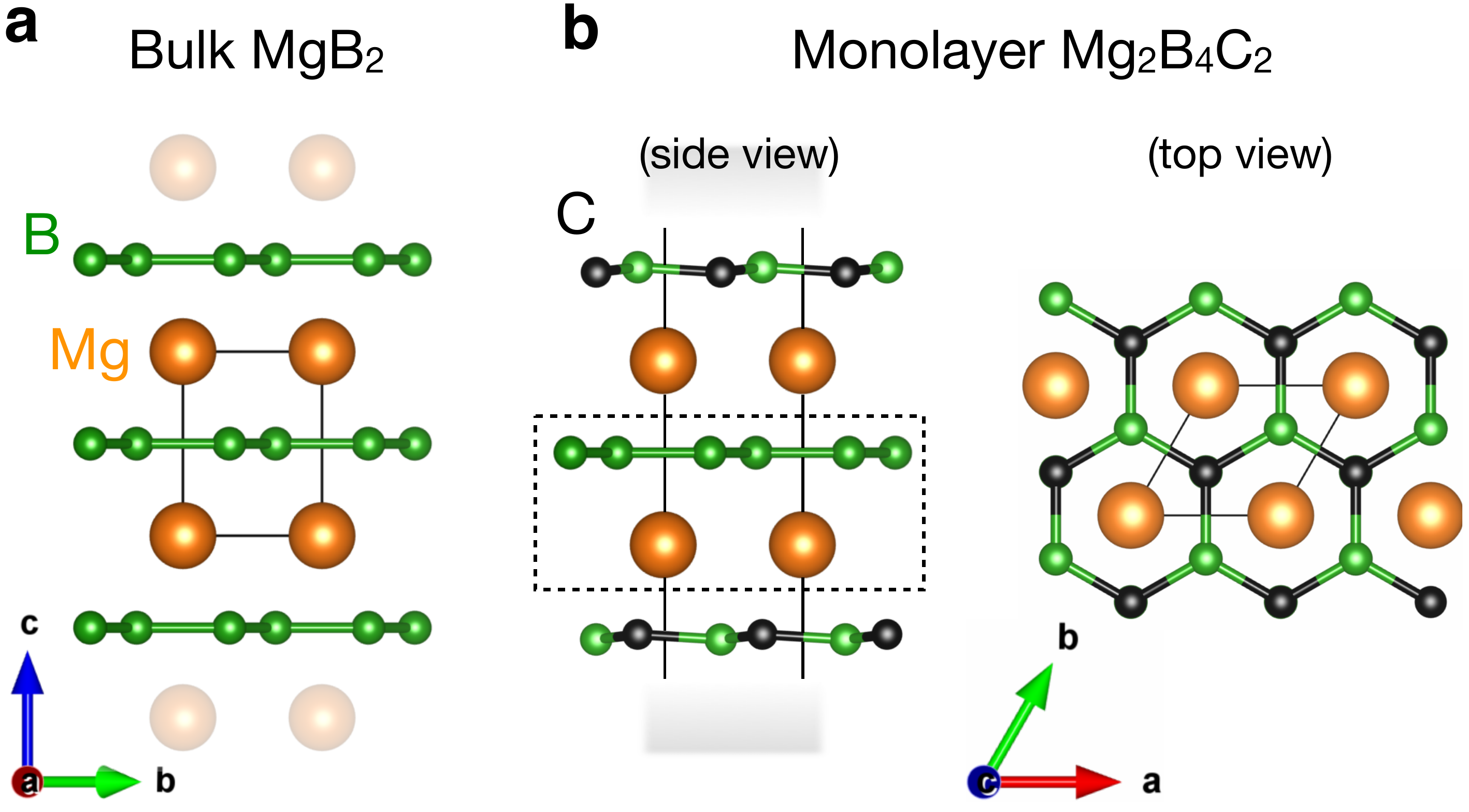}
\caption{Crystal structure. ($\bf{a}$) bulk MgB$_2$, and 
($\bf{b}$) side (as viewed from $\vec{\bf{a}}$) and top views of monolayer 
Mg$_2$B$_4$C$_2$ (Mg: orange, B: green, C: black).  
Solid black lines mark the unit cell boundaries, and shaded grey 
areas represent vacuum in the left panel of ($\bf{b}$). 
The region marked by dashed black lines in ($\bf{b}$) can be 
arbitrarily repeated (see text). }  
\label{fig:struct}
\end{figure*}

{\bf{Material design strategy.}} 
We start by describing our rationale for design of 
a stable MgB$_2$-like 2D superconductor having inert surfaces. 
Generally, layered vdW materials can be exfoliated to produce their 2D
analogues~\cite{geim2013van}. Although bulk MgB$_2$ has a layered structure, 
it is not a vdW material. Bulk MgB$_2$ crystallizes in space group $P_{6}/mmm$ (\#191) 
containing alternating layers of Mg and B 
atoms stacked along the $\vec{c}$ lattice direction, 
as shown in Fig.~\ref{fig:struct}(a)~\cite{XXXi2008}. 
The bonding between the Mg and B atoms is purely ionic, which means that Mg atoms 
donate two electrons to B atoms, thereby making each Mg $2+$ and each B $1-$. 
Since a B$^{-}$ is isoelectronic to a charge neutral carbon atom, 
a B-B sheet is structurally analogous to a single layer graphene, 
but it has a different ordering of bands than those of graphene. 
A simple exfoliation of MgB$_2$ into a 2D slab with B (or Mg) termination  
would yield a highly reactive electron-rich (or hole-rich) surface layer 
that is chemically unstable. 
    
\ssrev{We propose that one can passivate the charged surface layers in the 
MgB$_2$ slab by systematically substituting one boron by one carbon atom at the top and bottom surfaces of the slab. 
Fig.~\ref{fig:struct}(b) shows the top and side views of a Mg$_2$B$_4$C$_2$ monolayer 
designed using the aforementioned strategy. }
Strikingly, we find that modestly repeating the intermediate Mg-B layers, 
$i.e.$, the layers sandwiched between the top and bottom surfaces 
(highlighted using dashed rectangle in Fig.~\ref{fig:struct}(b)), 
thereby making thicker slabs of (MgB$_2$)$_n$C$_2$ 
while remaining in the quasi-2D limit, 
$n$ being the total number of Mg layers, retains 
the key features of the Mg$_2$B$_4$C$_2$ monolayer.
The electronic bandstructures calculated up to $n=5$ are 
shown in the Supplementary Material (SM)~\cite{SM}. 
This feature could be particularly useful in the experimental 
realization of 2D superconductivity in Mg$_2$B$_4$C$_2$. 
\ss{We note that MgB$_2$ monolayer can also be passivated 
by an appropriate hydrogenation process~\cite{Bekaert2019}. }

Mg$_2$B$_4$C$_2$ monolayer, shown in Fig.~\ref{fig:struct}(b), belongs to the layer group 
$p\bar{3}m1$ (\#72) having DFT (PBE) optimized lattice parameters $a=b=$ 2.87\,\AA. 
The absolute thickness between the top and bottom atomic layers is 7.14\,\AA, 
whereas, the interlayer spacing between the adjacent Mg and B-B, and Mg and B-C (C-B) 
layers is $\sim$1.8\,\AA,~and $\sim$1.7\,\AA, respectively. 
We note that the inversion symmetry is preserved due to the inverted ordering of the top and bottom layers 
in the structure shown in Fig.~\ref{fig:struct}(b). 
However, one could break the inversion symmetry by replicating the top and bottom layers, 
$i.e.,$ by making the top and bottom layers alike, either both as B-C or both as C-B. 
Our calculations suggest that the structure with inversion symmetry is energetically more favorable 
(5\,meV/f.u.) than the structure with broken inversion symmetry; 
although both structures are dynamically, elastically, and mechanically stable
since they exhibit all positive phonon frequencies, positive elastic constants, and 
satisfy the Born-Huang mechanical stability criteria (see SM~\cite{SM}). 
The only qualitative difference in the electronic properties of the structure with broken  
inversion symmetry is a small lifting of some band degeneracies at the $K$ 
high symmetry point (see SM~\cite{SM}). 
This effect is analogous to the application of a perpendicular electric field to a 
bilayer graphene~\cite{li2012marginal}. 

In this article, hereafter, we focus only on \ss{ground state structure of} a monolayer  
Mg$_2$B$_4$C$_2$ with preserved inversion symmetry. 
\ss{We note, all other possible atomic configurations of this composition are 
higher in energy (see SM~\cite{SM}). Furthermore, our exfoliation energy calculations 
(see Table S2~\cite{SM}) suggest that the reported monolayer Mg$_2$B$_4$C$_2$ 
belongs to the ``easily exfoliable" category, as classified by 
Mounet et al.~\cite{Mounet2018}. }

\begin{figure*}[ht]
\centering \includegraphics[width=2.0\columnwidth]{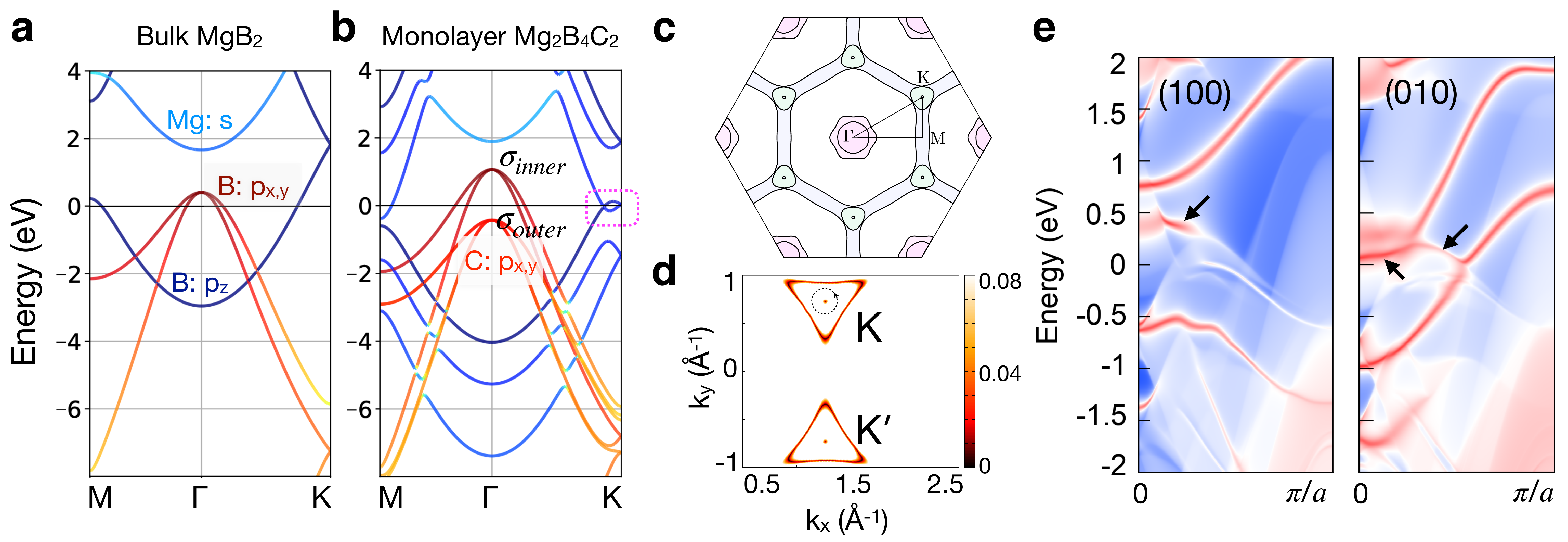}
\caption{Atomic orbitals projected electronic band structure of ($\bf{a}$) bulk MgB$_2$, 
and ($\bf{b}$) monolayer Mg$_2$B$_4$C$_2$ calculated without spin-orbit coupling (SOC) 
along the high symmetry direction of BZ. Cyan, red, and blue colors represent the 
contribution from the s, $p_{x,y}$, and $p_{z}$ orbitals, respectively. 
($\bf{c}$) Calculated Fermi surface of monolayer Mg$_2$B$_4$C$_2$. 
Light pink/green, and grey colors depict hole/electron, 
and intertwined electron-hole pockets, respectively. 
($\bf{d}$) Energy bandgap ($E_{gap}$) plotted in color scale (eV units) 
in the vicinity of a K high-symmetry point. 
\ssrev{The dashed circle marks the $k$-loop along which Berry phase was computed.} 
($\bf{e}$) The local electronic density of states of the (100) and (010) edge states spectrum. 
Red/White color denotes the states near the edge/interior of the 2D system. 
Topological nontrivial edge states are marked using arrows. }
\label{fig:bands}
\end{figure*}


\vspace{0.5cm} 
{\bf{Topological electronic properties of Mg$_2$B$_4$C$_2$ monolayer. }} 
After describing the crystal structure and its stability, 
we now focus on the topological electronic properties of Mg$_2$B$_4$C$_2$ monolayer. 
We begin by summarizing the key features of the electronic
structure of bulk MgB$_2$~\cite{XXXi2008} 
from which Mg$_2$B$_4$C$_2$ monolayer is derived. 
As shown in Fig.~\ref{fig:bands}(a), the Fermi surface of MgB$_2$ is composed of 
boron $p$ orbitals, where $p_{x,y}$ orbitals hybridize with $s$ orbitals to form strong covalent 
in-plane $\sigma$ bonds at the zone center, while the unhybridized $p_{z}$ orbitals form relatively weak 
out-of-plane $\pi$ bonds at zone boundaries (Mg acts as electron donor). Due to such a distinct 
Fermi-surface geometry, two  superconducting gaps exists in bulk MgB$_2$: 
(i) the stronger $\sigma$ gap of $\sim$7\,meV, and 
(ii) the weaker $\pi$ gap of $\sim$2-3\,meV~\cite{Szab_PRL2001, ChoiNature2002, Choi_PRB2002, Iavarone02, Iavarone_SciTech2002, MarginePRB2013, DMouPRB2015, AperisPRB2015, Kortus01}.  
Different symmetries of the $\sigma$ and $\pi$ bonds largely suppress the impurity
scattering in MgB$_2$~\cite{Iavarone02, XXXi2008, MAZIN200349, Kortus01}. 

Since the basic structure and charge neutrality of MgB$_2$ is
preserved in monolayer Mg$_2$B$_4$C$_2$, the electronic band structure 
of monolayer Mg$_2$B$_4$C$_2$ qualitatively resembles with that of the bulk MgB$_2$, 
as shown in Fig.~\ref{fig:bands}(a,b), but with some additional features. 
For instance, there is a new set of degenerate $\sigma$ bands ($\sigma_{outer}$) 
present at $\Gamma$ below the Fermi level arising from the $p_{x,y}$ orbitals of 
the outer boron-carbon layers. 
The other set of degenerate $\sigma$ bands ($\sigma_{inner}$) at $\Gamma$ 
that cross the Fermi level (also present in MgB$_2$) 
are formed by the $p_{x,y}$ orbitals of the inner boron-boron layer. 
These two sets of $\sigma$ bands are almost parallel and split by 
$\sim1.6$\,eV at $\Gamma$. Since the $\sigma_{outer}$ bands are 
completely occupied, they should, in principle, have no contribution in 
superconductivity, unless there is a large external field applied in 
a FET-like geometry~\cite{Ye1193}.

In addition to a new set of $\sigma$ bands at $\Gamma$, we notice 
the presence of Dirac-like band crossings at the K point, 
as well as along the high-symmetry directions near the K point of 
monolayer Mg$_2$B$_4$C$_2$. 
Regardless of their topological nature, these band crossings 
\ssrev{at the Fermi level, highlighted using a dashed magenta box in Fig.~\ref{fig:bands}(b), 
yield a large DOS at the Fermi level (almost 30\% larger than in bulk 
MgB$_2$), which contributes substantially to the total el-ph coupling in the studied monolayer. 
We note that the Dirac-like crossing at K is also present in bulk MgB$_2$, 
but it is situated well-above the Fermi level~\cite{Jin_npj2019}. }
The Dirac-like band crossings in Mg$_2$B$_4$C$_2$ monolayer are 
formed by highly dispersing $p_{z}$ orbitals of carbon and boron atoms 
(see SM~\cite{SM} for details). 
Thus, the Fermi surface of Mg$_2$B$_4$C$_2$ monolayer, 
shown in Fig.~\ref{fig:bands}(c), embodies three main features: 
(i) two hole pockets at $\Gamma$ (one circular and another that 
takes the shape of the BZ) composed of $\sigma$ bonded 
boron $p_{x,y}$ orbitals, 
(ii) an electron pocket at M formed by boron $p_{z}$ orbitals, 
and (iii) intertwined electron and hole pockets at the K point and 
along K--M high-symmetry line, formed by $\pi$ bonded carbon and boron $p_{z}$ orbitals. 
We note that all these pockets show very strong coupling to the phonon modes, and,  
as a result, they play the key role in governing superconductivity 
in Mg$_2$B$_4$C$_2$ monolayer, as we discuss later. 
Furthermore, the sharp and well-defined (almost flat) boundaries of 
the charge-carrier pockets at the Fermi surface 
set up the stage for the possible realization of Kohn-like divergencies~\cite{WKohn1959}, 
and charge-density wave ordering~\cite{CDW_RMP1988, ZhuPNAS2015} 
in this 2D system, which is beyond the scope of present work 
and calls for a more comprehensive attention in the future.  

By plotting the energy bandgap (E$_{gap}$) distribution in the vicinity of the K points, 
we discover presence of a triangular nodal line in the vicinity of each K point, 
as shown in Fig.~\ref{fig:bands}(d). However, this is not a truly gapless nodal line 
since a small E$_{gap}$ ($\sim$5\,meV) exists due to the subtle breaking of   
$M_{z}$ mirror symmetry. 
\ssrev{It is worth noting that the Dirac point at K is protected by 
the $C_{3v}$ rotation, inversion, and time-reversal symmetries; }
a small gap opens at Dirac points when the inversion symmetry is broken by 
making the top and bottom B-C layers identical~\cite{SM, GibsonPRB2015}.  
Although there are theoretical proposals suggesting the possibility of 
topological superconductivity in Dirac semimetals~\cite{Kobayashi_PRL2015}, 
we think that the so-far studied models are quite simple, and 
this topic requires a more thorough examination before any exotic
effects can be confidently claimed here.

In order to prove the nontrivial topological nature of Dirac points, 
we compute the Berry phase along a $k$-loop enclosing the
gapless point at K\ssrev{, as marked using dashed lines in Fig.~\ref{fig:bands}(d).}  
Our calculations yield a nontrivial Berry phase of
$\pi$ for Dirac points at K. 
\ssrev{We note, this exercise could not be performed for the Dirac nodal 
line near K because enclosing the nodal line residing in the $k_{x}$-$k_{y}$ plane 
would required a $k$-loop encircling along $k_{z}$ and $k_{z}$ is not defined for a 2D system.  
Nevertheless, the presence of time-reversal and spatial-inversion symmetries of 
Mg$_2$B$_4$C$_2$ monolayer enables us to determine the $Z_{2}$ topological invariants 
using the Fu-Kane criterion~~\cite{LiangFu2007}. }
The inversion parity eigenvalues of the electronic wavefunction of all 12 occupied bands 
at four time-reversal invariant momenta (TRIM) points are given in
Table~\ref{tab:parity}. The product of all parity eigenvalues
($\delta$) at each TRIM is also listed in Table~\ref{tab:parity}. We
find that the $Z_{2}$ topological index is nontrivial due to $\delta$
= -1 at three TRIM points. 
Here, we note that bulk MgB$_2$ has a weak $Z_{2}$ topological index (0;\,001) 
due to the band-inversions occurring at the $\Gamma$ and $A$  (0, 0, 0.5) 
high-symmetry points of 3D hexagonal BZ~\cite{Jin_npj2019}. 
\ssrev{Robust topological surface states have recently been experimentally 
observed in bulk MgB$_2$~\cite{XZhouPRB2019}. }

\begin{table}[h!]
\renewcommand{\arraystretch}{1.3}
\caption{Parity eigenvalues of all occupied bands and their products
  at four TRIM points \\ }
\label{tab:parity}
\begin{tabular}{ | c |  c  | c |} 
\hline
 TRIM               & Parity eigenvalues & $\delta$ \\
\hline
$\Gamma$ (0, 0, 0)   	 & $+-+-+-++--++$     &   $-1$    \\
$M_1$ (0.5, 0.0, 0.0)	 	& $-+-+-++-+---$      &     $-1$  \\
$M_2$ (0.0, 0.5, 0.0) 	 & $-+-+-++-+---$     &     $-1$  \\
$M_3$ (0.5, 0.5, 0.0)  	 & $-+-+-++-+--+$     &    $+1$  \\
\hline
\end{tabular}
\end{table}

Since the nontrivial topology in 2D systems is often manifested in the gapless 1D
edge states, we further confirm the nontrivial topological features of monolayer 
Mg$_2$B$_4$C$_2$ by computing the local density of states at (100) and (010) edges
of 60 unit cell thick nano-ribbons. 
Topologically nontrivial 1D edge states connecting band-crossing points 
were obtained at both (100) and (010) edges, as shown in Fig.~\ref{fig:bands}(e), 
thus, proving the nontrivial topology of the Mg$_2$B$_4$C$_2$ monolayer.


\vspace{0.5cm}

\begin{figure*}[ht!]
\centering \includegraphics[width=2\columnwidth]{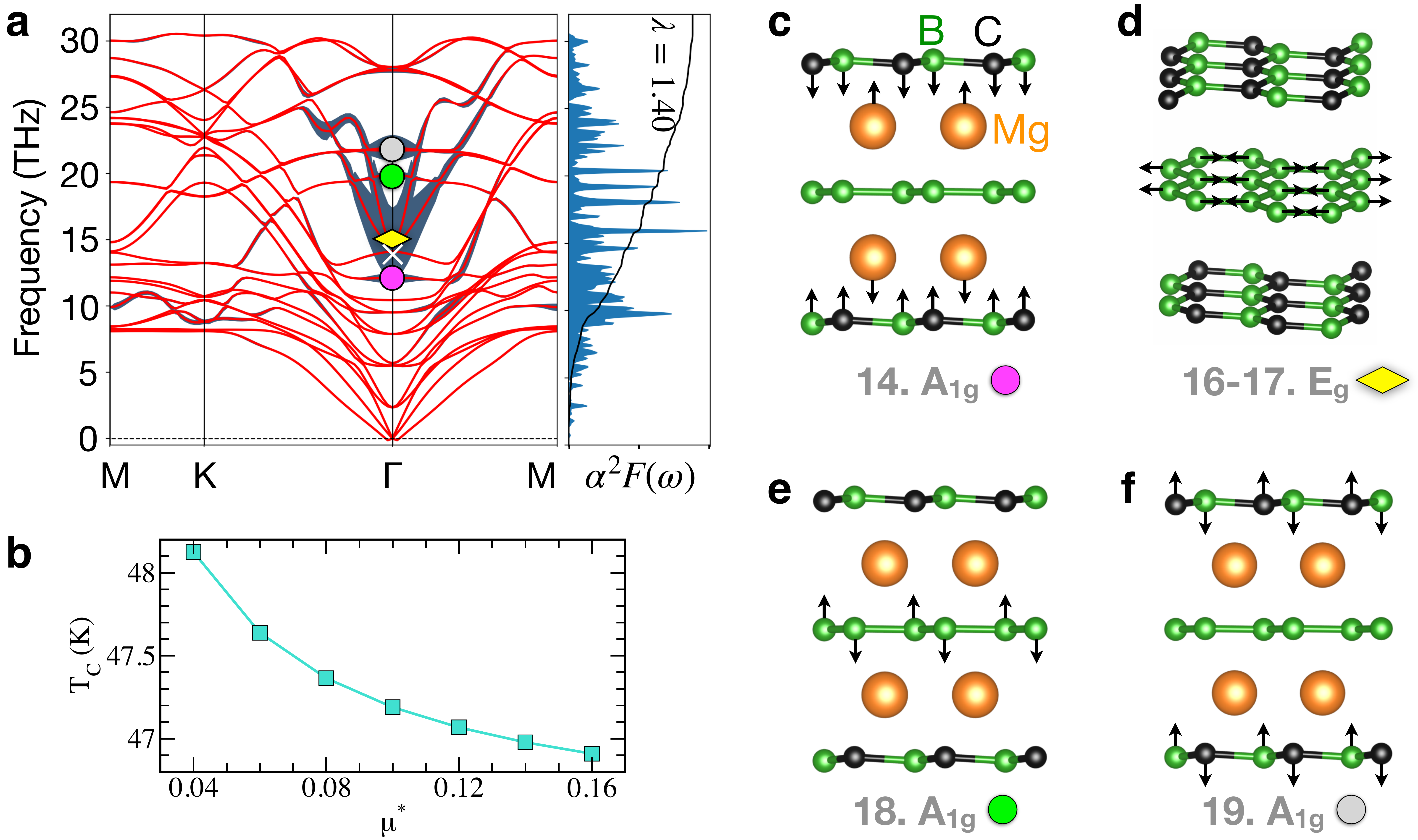}
\caption{
 ($\bf{a}$) Calculated phonon spectrum of Mg$_2$B$_4$C$_2$ monolayer 
  with phonon linewidths $\lambda(\mathbf{q},n)$ plotted using shaded blue color. 
  To avoid large overlap of $\lambda(\mathbf{q},n)$ with the phonon spectra, 
  we have divided the intensity by a factor of two. 
 \ss{The colored circles mark the three out-of-plane nondegenerate $A_{1g}$ modes (indices 14, 18, and 19), and the yellow diamond marks one in-plane doubly degenerate $E_{g}$ mode at $\Gamma$. These modes exhibit dominant el-ph coupling.  The atomic displacement patterns corresponding to these modes are shown in ($\bf{c-f}$). 
 The nondegenerate $A_{1u}$ mode (index 15) marked using symbol `$\times$' does not contribute to the total el-ph coupling, although it appears to be buried in the large $\lambda(\mathbf{q},n)$ of the $E_{g}$ mode. 
 The numerals 14, 15, 16, 17, 18, and 19 denote the phonon mode index as counted 
 from the lowest to the highest frequency modes ($i.e.,$ 1-3 for acoustic modes). 
 Mg atoms are omitted in ($\bf{c}$) for the sake of clarity.  
 The Eliashberg spectral function $\alpha^{2}F(\omega)$ along with the el-ph coupling 
 constant $\lambda$ in plotted in the right panel of ($\bf{a}$).
  ($\bf{b}$) Estimated $T_c$ as a function of the $\mu^*$ parameter. }}
\label{fig:elph}
\end{figure*}

{\bf{Electron-phonon coupling and superconductivity in Mg$_2$B$_4$C$_2$. }} 
We find that the roots of superconductivity in Mg$_2$B$_4$C$_2$ monolayer 
are same as in bulk MgB$_2$~\cite{Pickett_PRL2001, bohnen2001phonon, Iavarone02, 
ChoiNature2002, Choi_PRB2002, Pickett_Nature2002, MAZIN200349, XXXi2008, 
PICKETT2008, Pickett_design2006}. 
However, the main advantageous factor in Mg$_2$B$_4$C$_2$ is that, in addition to 
the doubly degenerate $E_{2g}$ modes that govern superconductivity in MgB$_2$, 
numerous other phonon modes strongly couple to the electronic states near the 
Fermi level yielding a much larger overall el-ph coupling, and thus, resulting in 
a considerably higher $T_c$.

The calculated phonon spectrum of Mg$_2$B$_4$C$_2$ monolayer, 
shown in Fig.~\ref{fig:elph}(a), contains a total of 24 phonon modes 
(8 atoms/cell) having the following mode symmetry at $\Gamma$: 
\begin{equation}
\begin{aligned}
\Gamma\textsubscript{acoustic} = \emph{A}\textsubscript{2u} \oplus \emph{E}\textsubscript{u}, ~\text{and} \\ 
\Gamma\textsubscript{optic} = 4\emph{A}\textsubscript{1g} \oplus 3\emph{A}\textsubscript{2u} \oplus 3\emph{E}\textsubscript{u} \oplus 4 \emph{E}\textsubscript{g}. 
\end{aligned}
\end{equation}
Here, $\emph{A}\textsubscript{1g}$ and $\emph{E}\textsubscript{g}$ 
are Raman-active modes, whereas, $\emph{A}\textsubscript{2u}$ and 
$\emph{E}\textsubscript{u}$ are infrared-active modes. 
\ss{In Fig.~\ref{fig:elph}(c-f), we show the atomic vibration 
patterns for the four phonon modes, 
namely, three nondegenerate $\emph{A}\textsubscript{1g}$ 
modes (index 14, 18, and 19) and one degenerate $\emph{E}\textsubscript{g}$ 
mode (indices 16-17), which exhibit the dominant el-ph coupling.  
All these $\emph{A}\textsubscript{1g}$ modes correspond to 
the out-of-plane vibrations of the Mg, inner B-B, and outer B-C layers, 
while the $\emph{E}\textsubscript{g}$ mode corresponds to 
the in-plane stretching of the inner B-B layer. }
The $\emph{A}\textsubscript{1g}$ modes primarily  
modulate the el-ph coupling associated with the $\pi$ 
bonded $p{_z}$ orbitals contributing to the electron and 
hole pockets located at the BZ boundaries. 
Whereas, the doubly degenerate $\emph{E}\textsubscript{g}$ 
mode couples with the $\sigma$ bonded 
$p_{x,y}$ orbitals forming the hole pockets located at $\Gamma$. 
Here, it is worth noting that the higher frequency 
$\emph{E}\textsubscript{g}$ modes (indices 21-22) that 
correspond to the in-plane stretching of the 
outer B-C layers do not make a significant contribution 
to the overall el-ph in this system, which is as expected 
since these modes modulate the occupied $\sigma_{outer}$ bands 
located well-below the Fermi level at $\Gamma$ [see Fig.~\ref{fig:bands}(b)]. 
\ss{However, these modes may participate in the superconductivity 
when the system is doped with $p$-type charge carriers~\cite{Bekaert2019}. }

Since the electronic and vibrational band structures of inner B-B 
and outer B-C layers are essentially independent of each other, 
we predicate that the reported properties of the studied 
Mg$_2$B$_4$C$_2$ monolayer would be retained
even when the number of the inner B-B layers are repeated (until a critical thickness), 
thus making the system thicker. This feature might greatly simplify 
the eventual realization of superconductivity in Mg$_2$B$_4$C$_2$. 

To quantify the superconducting properties of Mg$_2$B$_4$C$_2$ monolayer, 
\ssrev{we employ the McMillian-Allen-Dynes theory derived from the isotropic Migdal-Eliashberg 
formalism~\cite{Migdal1958, Eliashberg60, PBAllen1983}}
which relies on the calculation of the el-ph coupling matrix elements within DFT. 
The calculated matrix elements correspond to the transition probabilities 
of different Kohn-Sham states induced by a change in the potential 
due to a small ionic displacement. 
Thus, these matrix elements provide the main ingredients 
to calculate the el-ph coupling strength 
and the Eliashberg spectral function $\alpha^{2}F(\omega)$ 
as a function of the phonon frequency $\omega$. 
Since the physical process behind the phonon-mediated 
superconductivity is the exchange of a phonon between two electrons, 
a strong el-ph coupling is desired to achieve a high-$T_c$ in a BCS superconductor. 
Theoretical details of such calculations are explained in numerous other 
papers~\cite{giustino2017electron, giustino2007electron, MarginePRB2013}. 

In Fig.~\ref{fig:elph}(a), we plot the calculated phonon linewidth 
$\lambda({\bf{q}}, n)$ for each phonon mode $n$ at each wave 
vector $\bf{q}$ using blue color. 
Note that the plotted phonon linewidth is scaled down by a 
factor of two to avoid large overlap 
with the neighboring phonon branches. 
\ss{The largest contribution to the total el-ph coupling strength 
comes from three nondegenerate $\emph{A}\textsubscript{1g}$ modes   
and one doubly degenerate $\emph{E}\textsubscript{g}$ mode,  
as marked in Fig.~\ref{fig:elph}(a). 
We note that the $\emph{A}\textsubscript{2u}$ mode 
(index 15), marked using `$\times$ in Fig.~\ref{fig:elph}(a), 
does not contribute to the total el-ph coupling, although it appears  
buried in the large $\lambda(\mathbf{q},n)$ overlap from the $E_{g}$ mode. }
Notably, in addition to the aforementioned $\emph{A}\textsubscript{1g}$ and 
$\emph{E}\textsubscript{g}$ phonon modes, various other modes make 
relatively smaller contributions to the overall el-ph coupling strength, 
as revealed by the Eliashberg spectral function $\alpha^{2}F(\omega)$ 
plot shown in the right panel of Fig.~\ref{fig:elph}(a). 

In addition to the el-ph coupling, the net \ssrev{phonon linewidth $\lambda({\bf{q}}, n)$ can 
have some contribution from the phonon-phonon (ph-ph) interactions owing 
to the phonon anharmonicity~\cite{Choi_PRB2002}}. Therefore, we thoroughly investigate ph-ph 
interactions by computing ph-ph linewidth using the {\it ab-initio} 
molecular dynamics simulations. 
%
%
In this approach, we mapped the forces, 
obtained from the finite-temperature molecular dynamics simulations, 
evaluated in a 3 $\times$ 3 $\times$ 1 supercell onto 
a model Hamiltonian describing the lattice dynamics. 
This temperature dependent effective potential (TDEP) 
technique~\cite{hellman2013temperature_b, hellman2013temperature_a} 
enabled us to calculate the third-order response 
from the effective renormalized interatomic force constants. 
Our calculations revealed that the ph-ph linewidths are 
an order of magnitude smaller than the el-ph linewidths. 
The maximum value of obtained ph-ph linewidth is $\sim$2\,meV, 
which is much smaller compared to the 
el-ph linewidth values that are typically larger than 
$\sim$70\,meV in the studied system.  
This result implies that, although the system inherits 
some anharmonic effects, we can safely discard  
the ph-ph contributions in the study of its superconducting properties.

Based on the BCS theory of superconductivity and 
above results, we estimate the critical temperature $T_c$ using the 
McMillian-Allen-Dynes formula~\cite{BCS1957, McMillan1968, AllenDynes1975}: 
\begin{equation}
\begin{aligned}
T_c=\frac{\omega_{log}}{1.2} exp \left[
-\frac{1.04(1+\lambda)}{\lambda-\mu^{*}(1+0.62\lambda)}\right], 
\end{aligned}
\end{equation}
where $\omega_{log}$ is the logarithmic averaged phonon frequency, 
$\lambda$ is the total el-ph coupling constant, and   
$\mu^{*}$ is the effective screened Coulomb repulsion 
constant 
\ssrev{with a typical value ranging 
from 0.04--0.16 (see Table~\ref{tab:SC})~\cite{bohnen2001phonon, ChoiNature2002, Choi_PRB2002, Choi09, dai12}. }
We obtain $\lambda$ by integrating the cumulative 
frequency-dependent el-ph coupling $\lambda(\omega)$ 
given by the following expression: 
\begin{equation}
\begin{aligned}
\lambda(\omega)=2 \int_{0}^{\omega} \frac{\alpha^{2} F(\omega)}{\omega} d\omega 
\end{aligned}
\end{equation}
We find a fairly large value of $\lambda = 1.40$, 
which is considerably larger than the one reported for 
bulk MgB$_2$ ($\lambda_{bulk} = 0.73$~\cite{bohnen2001phonon}, 
and 0.61~\cite{Choi_PRB2002}). 
\ss{We observe that the estimated $T_c$ does not vary drastically as a function 
of $\mu^*$, as shown in Fig.~\ref{fig:elph}(b). This is consistent with an earlier work 
by Choi et al.~\cite{Choi_PRB2002}, which reported that the superconducting 
properties of MgB$_2$ are not very sensitive to the $\mu^*$ parameter 
within the isotropic McMillian-Allen-Dynes formalism. }
 We note that for bulk MgB$_2$, $\mu^*$ = 0.05 
has been used to get the correct estimate of 
$T_c$ $\sim$ 40\,K~\cite{bohnen2001phonon}. 
Therefore, using the 
\ssrev{McMillian-Allen-Dynes formula~\cite{BCS1957, McMillan1968, AllenDynes1975},}
we estimate the $T_c$ of Mg$_2$B$_4$C$_2$ monolayer 
to be in the range 47--48\,K without any doping or strain. 
\ss{Our results are consistent with a recent study~\cite{Bekaert2019} in which 
$T_c = 67\,K$ and $\lambda = 1.46$ was predicted in a 
hydrogenated MgB$_2$ monolayer by solving the fully anisotropic Eliasberg equations. 
We argue that the predicted $T_c$ in Mg$_2$B$_4$C$_2$ monolayer can be 
further enhanced by biaxial strain~\cite{Bekaert2019, YanLuo2020} or by 
p-doping~\cite{Bekaert2019}.  }
In passing, we would like to mention that the predicted $T_c$ 
could moderately vary if a \ss{fully anisotropic Migdal-Eliashberg theory~\cite{Choi_PRB2002, 
MarginePRB2013, JJZhengPRB2016, Bekaert2019} or SC-DFT~\cite{SCDFT_PRB2005, 
SCDFT2_PRB2005, SannaPRB2007, SannaPRL2020} }
is employed. 
This is particularly important here because the applicability of the 
McMillian-Allen-Dynes formula becomes limited in the case of large el-ph coupling.

\begin{table*}[htb!]
    \renewcommand{\arraystretch}{1.30}
\caption{Listing of superconducting parameters required for the prediction of $T_c$ using the McMillian-Allen-Dynes formula for some reported 2D phonon-mediated superconductors 
(data for bulk MgB$_2$ is included for comparison). 
This table includes data of effective Coulomb screening parameter $\mu^{*}$, 
 electronic DOS at the Fermi level $N$(E$_F$) (in states/spin/Ry/cell), 
 logarithmic averaged phonon frequency $\omega_{log}$ (in K), 
 total electron-phonon coupling constant $\lambda$, and 
 estimated $T_c$ (in K). Experimental $T_c$ values are noted in the table. \\ }
\label{tab:SC}
\begin{tabular}{ | l  | c |  c | c | c | c | c | } 
\hline
 Compounds   & $\mu^{*}$ &  $N$(E$_F$)  &  $\omega_{log}$  &  $\lambda$   &   $T_c$   &   Ref.  \\
\hline
B$_2$C	  	 &	0.10  	&	 	 & 	315  &	0.92  &	19  &	 \cite{dai12}	\\
CaC$_6$ 	  &	0.115   	&	 	 & 	446   &	0.40  &	1.4  &	 \cite{GianniNatPhys2012}	\\
LiC$_6$ 	  &	0.115   	&	 	 & 	400   &	0.61  &	8.1  &	 \cite{GianniNatPhys2012}	\\
LiC$_6$ 	  &	    	&	 	 &    &	0.58 $\pm$ 0.05  &	5.9 [Exp.]  &	 \cite{Ludbrook_PNAS2015}	\\
LiC$_6$ 	  &	 0.12/0.14/0.16   &	 	 &    & 0.55 & 7.6/5.9/5.1 &	 \cite{JJZhengPRB2016}	\\
2H-NbSe$_2$ 	  &	   	&	 	 & 	    &	0.75  &	3.1 [Exp.] &	 \cite{Xiaoxiang2015}	\\
2H-NbSe$_2$ 	  &	 0.15, 0.16  	&	 	 & 	134, 145     &	0.84, 0.67  &	4.5, 2.7  &	 \cite{LianNanoLett2018, FeipengPRB2019}	\\
C$_6$CaC$_6$ 	  &	   	&	 	 & 	    &	   &	4.0 [Exp.] &	 \cite{IchinokuraACSNano2016}	\\
C$_6$CaC$_6$ 	  &	0.207/0.155   	&	 	 & 	    &	   &	6.8/8.1  &	 \cite{ERMargine2016, MazinBalatsky2010, jishi2011theoretical}	\\
B$_2$O	  	 &	0.10  	&	5.4 	 & 	250  &	0.75  &	10.3  &	 \cite{YanLuo2020} 	\\
LiBC	  	 &	0.13 	&	10.9 	 & 	   &	0.59  &	65  &	 \cite{modakPRB2021} 	\\
bulk MgB$_2$	  	 &	0.05  	&	9.8  	 & 	707   &	0.73  &	40  &	 \cite{bohnen2001phonon} 	\\
bulk MgB$_2$	  	 &	0.13  	&	9.8 	 & 	  	 &	0.61  &	39  &	 \cite{Bekaert2017} 	\\
monolayer MgB$_2$	  	 &	0.13  	&	13.1 	 & 	   &	0.68  &	20  &	 \cite{Bekaert2017} 	\\
monolayer H-MgB$_2$	  &	0.13  	&	19.2 	 & 	   &	1.46  &	67  &	 \cite{Bekaert2019} 	\\
\hline
\multirow{3}{*}{Mg$_2$B$_4$C$_2$} &	0.04  &  \multirow{3}{*}{12.6}	 & 	\multirow{3}{*}{506}  &	\multirow{3}{*}{1.40}  &	48.1  &  Our work 	\\
 							 &	0.10  &   					 & 				 	&					&	47.2  &  Our work	 \\
 							 &	0.14  &   					 & 				 	&					&	47.0  &  Our work	 \\
\hline
\end{tabular}
\end{table*}

In order to highlight the novelty of our results, in Table~\ref{tab:SC} 
we list the theoretical superconducting parameters along with the 
estimated $T_c$ for some reported 2D phonon-mediated superconductors.  
The good agreement between the experimental data for 
LiC$_6$~\cite{GianniNatPhys2012, Ludbrook_PNAS2015, JJZhengPRB2016}, 
$2H$-NbSe$_2$~\cite{Xiaoxiang2015,HeilPRL2017,LianNanoLett2018, FeipengPRB2019}, and 
C$_6$CaC$_6$~\cite{IchinokuraACSNano2016,ERMargine2016} 
and the corresponding theoretical results obtained from the McMillian-Allen-Dynes theory 
further boosts our confidence in the predictive power of the employed theory.

\section*{Summary}
In summary, we present a 2D material Mg$_2$B$_4$C$_2$, similar to MgB$_2$, 
but with inert surfaces obtained by the replacement 
of outer B-B layers by B-C layers.
Our calculations suggest that this structure is 
dynamically, elastically and mechanically stable. 
It also features a nontrivial topological electronic band structure 
\ss{together with a large el-ph coupling ($\lambda=1.40$), which is 
more than twice as large as that of the bulk MgB$_2$ and comparable 
to that of in a hydrogenated monolayer MgB$_2$~\cite{Bekaert2019}. } 
Use of the \ssrev{standard McMillian-Allen-Dynes theory} predicts 
the superconducting transition temperature 
$T_c$ to be in the range of \ss{47--48\,K} without any doping or tuning of 
external parameters such as strain. 
To the best of our knowledge, this is \ss{among the} highest predicted intrinsic $T_c$ 
in a conventional BCS-type 2D superconductor to date. 
The studied material offers the two expected ingredients:  
(i) topological nontrivial electronic properties, and 
(ii) large intrinsic $T_c$, 
for practical realization of nontrivial topological superconductivity in 
2D~\cite{XQ_RMP2011}. In addition to the large el-ph coupling, 
the presence of sharp and well-defined flat boundaries of the 
charge-carrier pockets at the Fermi surface imply the possible 
realization of Kohn-like divergencies and charge-density wave 
ordering in this 2D system, which calls for a dedicated study in future.

\section*{Methods}

The electronic bands structure and phonon calculations were 
performed using density-functional theory (DFT) as implemented 
in the VASP package~\cite{vasp1, vasp2, vasp3, vasp4}. 
The phonopy~\cite{phonopy} and PyProcar~\cite{pyprocar} tools 
were used for the post-processing of data. 
The Perdew-Burke-Ernzerhof (PBE) exchange-correlation functional~\cite{pbe1996} and PAW
pseudo-potentials~\cite{Bloch94, vaspPaw} were used. 
The employed $k$-point grid for self-consistent calculations 
was $30 \times 30 \times 1$, and the cutoff for the
kinetic energy of plane waves was set to 700\,eV. 
A vacuum of thickness $\sim$30\,\AA~was added to 
avoid the periodic interactions along the $c$-axis. 
Since the spin-orbit coupling (SOC) effects were found to be negligible 
in the studied system, 
 SOC was not included in the reported calculations. 
The elastic and mechanical properties 
were analyzed using the MechElastic code~\cite{mechelastic, singh2021mechelastic}. 
The exfoliation energy was calculated using four different exchange-correlation approximations: 
the (PBE) GGA approximation~\cite{pbe1996}, 
the SCAN~\cite{scan} meta-GGA, vdW-DF2 GGA functional~\cite{vdWD2}, 
and SCAN together with the rVV10 correlation functional (SCAN+rVV10)~\cite{scanrVV10}. 
The topological properties of Mg$_2$B$_4$C$_2$ 
were studied by fitting the DFT calculated bandstructure 
to a real space tight-binding Hamiltonian obtained 
using the maximally localized Wannier functions (MLWFs) 
approach~\cite{mostofi2014updated, WU2018405}. 
The local density of states at (100) and (010) edges were 
calculated for 60 unit cells thick nano-ribbons using 
the WannierTools package~\cite{WU2018405} with 
vacuum added along the $c$-axis of the ribbon.

For the electron-phonon coupling matrix elements calculations, 
we used the abinit package~\cite{gonze2002first, gonze2005brief, 
gonze2009abinit, gonze2016recent, aldo2020}. 
We employed norm conserving pseudopotentials (using the ONCVPSP scheme of
Hamann~\cite{hamann2013optimized}), and a plane wave basis set up to
kinetic energies of 35\,Ha. Cell parameters were optimized by using the
PBE exchange-correlation functional as in VASP calculations. 
We used a uniform grid of \ss{$18 \times 18 \times 1$} for the ground state calculations,
and a phonon grid of \ss{$9 \times 9 \times 1$} for the phonon part. A total of \ss{288}
el-ph matrix elements were calculated. 
Calculations of the phonon interatomic force constants, 
and the el-ph coupling matrix elements
performed in this work used the second-order 
perturbation theory~\cite{baroni2001phonons, gonze1995adiabatic}.
The temperature dependent effective potential (TDEP) 
technique~\cite{hellman2013temperature_b, hellman2013temperature_a} 
was used to study the phonon-phonon interactions and phonon anharmonic effects.

\section*{Acknowledgements}
This work was supported in part by Fondecyt Grants No. 1191353 (F.M.), 3200697 (J.D.M.), 1180175 (V.E.), 1190361(E.M.),
Center for the Development of Nanoscience and Nanotechnology CEDENNA
AFB180001, and from Conicyt PIA/Anillo ACT192023.
S.S., K.R., and D.V. acknowledge the support from ONR 
Grants N00014-19-1-2073 and  N00014-16-1-2951. 
We also thank the NSF CAREER Award CHE-1351968 to A.N.A and 
DOE DE- SC0021375 project to A.H.R. 
This research was partially supported by the supercomputing
infrastructure of the NLHPC (ECM-02) and XSEDE which is supported by
National Science Foundation grant number ACI-1053575. The authors also
acknowledge the support from the Texas Advances Computer Center (with
the Stampede2 and Bridges supercomputers).

\bibliography{mg2b4c2_bib.bib}

\begin{thebibliography}{140}
\bibitem [{SM()}]{SM}%
  \BibitemOpen
  \href@noop {} {\enquote {\bibinfo {title} {See {S}upplemental {M}aterial
  ({SM}) at href for our results on the structure with broken inversion
  symmetry, thicker slabs, more details of the electronic band structure,
  tight-binding parameterization, exfoliation energy calculations, elastic
  properties, structural details, and phonons.}}\ }\BibitemShut {NoStop}%
\end{thebibliography}



\end{document}